\documentclass[a4paper]{article}

\usepackage{graphicx} 
\usepackage{hyperref}       %
\usepackage{url,color}            %
\usepackage{amsfonts}       %
\usepackage{amssymb,amsmath}
\usepackage{bbold}
\usepackage{subfig}
\usepackage{amssymb}
\usepackage{amsthm}
\usepackage{relsize} %
\usepackage{placeins} %
\usepackage{algorithm}

\usepackage[
backend=biber,
style=authoryear-comp,
sorting=ynt,
url=false,
doi=false,
isbn=false,
eprint=false,
natbib=true,
uniquename=false
]{biblatex}

\newcommand{\op}{\left(}
\newcommand{\cp}{\right)}
\newcommand{\ob}{\left[}
\newcommand{\cb}{\right]}
\newcommand{\R}{\mathbb{R}}
\newcommand{\E}{\mathbb{E}}

\newcommand{\1}{\mathbb{1}}
\newcommand{\X}{\mathcal{X}}
\newcommand{\ep}{\varepsilon}
\newcommand{\K}{\mathcal{K}}

\newcommand{\V}{\mathcal{V}}
\newcommand{\N}{\mathcal{N}}
\newcommand{\U}{\mathcal{U}}
\newcommand{\M}{\mathcal{M}}
\renewcommand{\P}{\mathcal{P}}
\renewcommand{\S}{\mathbb{S}}
\newcommand{\td}{\widetilde}

\newcommand{\QT}{\text{QT}}

\DeclareMathOperator{\tr}{tr}
\DeclareMathOperator{\conv}{conv}

\DeclareMathOperator{\diag}{diag}

\DeclareMathOperator{\argmin}{argmin}

\newcommand{\VVert}[1]{{\left\vert\kern-0.25ex\left\vert\kern-0.25ex\left\vert #1 
    \right\vert\kern-0.25ex\right\vert\kern-0.25ex\right\vert}}

\usepackage{appendix}
\usepackage{geometry}
\title{Variational Inference on the Boolean Hypercube\\ with the Quantum Entropy}

\author{ Eliot Beyler \and Francis Bach }

\date{Inria, Ecole Normale Supérieure,\\ PSL Research University
}

\begin{document}
\maketitle

\begin{abstract}
In this paper, we derive variational inference upper-bounds on the log-partition function of pairwise Markov random fields on the Boolean hypercube, based on quantum relaxations of the Kullback-Leibler divergence. We then propose an efficient algorithm to compute these bounds based on primal-dual optimization. An improvement of these bounds through the use of ``hierarchies,'' similar to sum-of-squares (SoS) hierarchies is proposed, and we present a greedy algorithm to select among these relaxations. We carry extensive numerical experiments and compare with state-of-the-art methods for this inference problem.
\end{abstract}

\section{INTRODUCTION}
Graphical models \citep[see, e.g.,][]{wainwrightGraphicalModelsExponential2007} are used to define structured relationships between random variables, making them useful in a variety of computational tasks such as bioinformatics \citep{durbinBiologicalSequenceAnalysis1998,pevznerComputationalMolecularBiology2000}, combinatorial optimization \citep{goemansImprovedApproximationAlgorithms1995}, constraint satisfaction \citep{manevaNewLookSurvey2007}, statistical physics \citep{landauStatisticalPhysicsVolume2013}, language and speech processing \citep{bleiLatentDirichletAllocation2003}, image processing \citep{blakeMarkovRandomFields2011}, or any other problems formulated as a probabilistic model.

\subsection{Log-partition function}
Undirected graphical models (a.k.a.~Markov random fields) define probability distributions up to an unknown normalization constant, called the log-partition function, which is notoriously hard to compute and a key quantity in inference tasks.

More precisely, we are given a probability $p$ on a finite set $\X$ (which we will take to be $\{-1,1\}^d$ in most of the paper) known to be proportional to $e^{f(x)}q(x)$ where $f:\X \rightarrow \R$ is a function and $q$ a base measure (often chosen to be the uniform measure). We want to find the normalizing constant $\Phi(f)$ such that:
$$
p(x) = e^{f(x) - \Phi(f)}q(x).
$$
Normalization of $p$ gives the following formula:
\begin{equation}
\label{eq:logp_logsum}
\Phi(f) = \log\sum_{x\in\X} e^{f(x)}q(x).
\end{equation}
Computing directly the sum (\ref{eq:logp_logsum}) is intractable in general, as the number of terms $|\X|$ will grow exponentially with the number of variables $d$ in our model. Our goal is to derive tractable approximations of this value, i.e., approximations that can be computed with polynomial complexity with respect to $d$. Moreover, we want to preserve properties of the log-partition function, such as its convexity in $f$.

\subsection{MRF on the Boolean hypercube}
The focus of our study is a special class of problems, namely pairwise Markov random fields (MRF) on the Boolean hypercube $\X = \{-1,1\}^d$, also known as the Ising model, meaning that $f$ is of the form:
$$
f(x) = \sum_{s\in V} \theta_s x_s + \sum_{(i,j)\in E} \theta_{ij} x_i x_j,
$$ 
where $G=(V,E)$ is an undirected graph, with vertices $V \subset \{1,\dots,d\}$. The structure of the graph can be used to characterize independence properties of the probability distribution \citep{wainwrightGraphicalModelsExponential2007}. When the structure does not matter, we can simply assume the graph to be complete by setting $\theta_{i} = 0$ if $i\notin V$ and $\theta_{ij} = 0$ if $(i,j)\notin E$.

When the underlying graph has a special tree structure, the computation of the log-partition function can be carried exactly with a message-passing algorithm, known as the sum-product or belief propagation algorithm (see for example \citet{kschischangFactorGraphsSumproduct2001}). However, this problem becomes intractable in the general case. As a consequence, research is focused on obtaining tractable approximation or bounds on the log-partition function.

\subsection{Kullback-Leibler divergence}
A powerful approach to finding and studying such approximation is variational inference \citep[see, e.g.,][and references therein]{wainwrightGraphicalModelsExponential2007}.
It relies on the Kullback-Leibler (KL) divergence.

For two probability measures $p$ and $q$ over a finite space~$\X$, we define the KL divergence between $p$ and $q$ as:
$$
D(p\Vert q) = \sum_{x\in\X} \log\op\frac{p(x)}{q(x)}\cp p(x).
$$
The KL divergence is non-negative, equal to zero if and only if $p=q$ and jointly convex in $p$ and $q$ \citep{coverElementsInformationTheory2005}.

\subsection{Variational inference}
The idea behind variational inference is to use convex duality (through Fenchel conjugates) to express the log-partition function as an optimization problem:
\begin{equation*}
\log\sum_{x\in\X} e^{f(x)}q(x) = \sup_{p \in \P(\X)} \sum_{x\in\X} f(x)p(x) - D(p\Vert q),
\end{equation*}
where $\P(\X)$ is the space of probability on $\X$. This equality is sometimes referred as the ``Donsker-Varadhan'' identity. One can then construct a tractable upper-bound (or lower-bound), by finding a tractable lower-bound (resp.~upper-bound) on the KL divergence, changing the optimization variable and have outer (resp.~inner) approximation of the optimization domain. Moreover, if the bound on the KL and the approximation of the optimization domain are convex, the resulting quantity is expressed as a convex optimization problem, thus can be computed efficiently.

In this framework, some methods have been developed to give upper-bounds on the log-partition function from convex lower bound on the KL\footnote{Note that other commonly used methods \citep[see, e.g.,][and references therein]{murphyProbabilisticMachineLearning2022,murphyProbabilisticMachineLearning2023} such as mean-field methods, have other properties (e.g., give lower bounds). These methods are not studied here.}, an important one being the tree-reweighted message passing (TRW) algorithm \citep{wainwrightNewClassUpper2005a}, another the log-determinant relaxation \citep{jordanSemidefiniteRelaxationsApproximate2003}. The TRW algorithm uses a bound on the entropy based on a combination of entropy for MRF on trees, for which we have a simple closed form formula. For the log-determinant relaxation, the entropy is bounded by the entropy of a Gaussian vector with similar covariance matrix.

\subsection{Challenges}
Recently, new links between sum-of-squares (SoS) optimization \citep{lasserreMomentsPositivePolynomials2010} and  quantum divergences between positive semi-definite matrices used as lower bounds on entropy have been made by \citet{bachInformationTheoryKernel2023,bachSumofSquaresRelaxationsInformation2024}.
\citet{bachInformationTheoryKernel2023,bachSumofSquaresRelaxationsInformation2024} also proposes to use these divergences in variational inference, but experiments are limited to the torus and lack actual comparisons with existing alternatives. Therefore, our work attempts to further explore the use of quantum divergences in machine learning by delving deeper into the particular case of the Boolean hypercube to try the answer the following questions:
can we develop practical methods of variational inference with these quantities? Can these methods be competitive with existing algorithms?

We answer these questions positively  by implementing efficient optimization algorithms for variational inference methods with the quantum entropy, which can be improved using hierarchies of relaxations similar to the ones used in SoS optimization and greedy selection between theses relaxations.

\subsection{Contributions}
In this paper, we make the following contributions:
\begin{itemize}
    \item We specialize the framework of \citet{bachInformationTheoryKernel2023,bachSumofSquaresRelaxationsInformation2024}, reviewed in section~\ref{sct:KL}, to the case of a pairwise MRF on the Boolean hypercube, and propose an efficient algorithm to compute an upper-bound on the log-partition function in section~\ref{sct:logp}, based on a primal-dual optimization algorithm.
    \item We propose in section~\ref{sct:impr} an improvement of bound with ``hierarchies,'' similar to SoS hierarchies, and present a greedy algorithm to select among these relaxations. 
    \item We carry in section~\ref{sct:exp} extensive numerical experiments and compare with the TRW and logdet relaxations.\footnote{Source code to reproduce the experiments is available at \url{https://github.com/ebeyler/variational_inference_quatum_entropy}}
\end{itemize}

\section{NOTATIONS}
We introduce the following notations:
\begin{itemize}
    \item $\X = \{-1,1\}^d$ the Boolean hypercube. We denote $\P(\X)$ the set of probability distributions on $\X$. For a function $f$ on $\X$ and a probability $p\in\P(\X)$, we write $\E_p[f(x)]= \sum_{x\in\X}f(x)p(x)$ the expectation of $f(x)$ under $p$.
    \item $\R$ the space of real numbers. For a positive integer $n$, $\R^n$ the space of vector of dimension $n$, $\M_n(\R)$ the space of square real matrices of dimension $n\times n$, $\S_n \subset \M_n(\R)$ the space the symmetric matrices,  $\S_n^+ \subset \S_n$ the positive semi-definite (PSD) matrices, and $\S_n^{++} \subset \S_n$ the positive definite matrices. For $A\in \S_n$, we write $A\succcurlyeq 0$ if $A\in \S_n^+$ and $A \preccurlyeq B$ if $B-A\succcurlyeq 0$.
    \item For a vector $y \in \R^n$ (or a matrix $A \in \M_n(\R)$), we write $y^T$ (resp. $A^T$) its transpose.
    \item For $A\in \M_n(\R)$, we write $\tr A = \sum_i A_{ii}$, $\diag(A) = (A_{ii})_{i=1...n}\in \R^n$ the diagonal of matrix $A$, and for $\lambda \in \R^n$, $\diag(\lambda)\in \M_n(\R)$ the diagonal matrix with diagonal $\lambda$.
    \item For $\lambda \in \R^n$, $\max(\lambda) = \lambda_i$ such that $\forall j \in \{1,...,n\}, \lambda_i \geq \lambda_j$.
    \item \sloppy For $h:\R\rightarrow\R$, and $A \in \S_n$, we write $h(A) = U^T\diag(h(\lambda_1),\dots,h(\lambda_n))U$ where $A = U^T \diag(\lambda_1,\dots,\lambda_n) U$ is the spectral decomposition of~$A$. The value $h(A)$ does not depend on the specific choice of orthogonal diagonalization basis and leads to a \emph{spectral function} $h:\S_n\rightarrow\S_n$.
\end{itemize}

\section{BOUNDING THE KL DIVERGENCE}
\label{sct:KL}
The function $f$ defined by $f(x) = \sum_{i} \theta_i x_i + \sum_{i,j} \theta_{ij} x_i x_j,$ can be rewritten as a quadratic form $f(x) = \varphi(x)^T F \varphi(x)$ with  $\varphi(x) = (1,x_1,\dots,x_d) \in \R^{d+1}$, and  $F$ a $(d+1)\times(d+1)$ matrix defined by $F_{1,i+1} = \theta_{i}$ and $F_{i+1,j+1} = F_{j+1,i+1} = \theta_{ij}/2$. Then we can write $f(x) = \tr \varphi(x)\varphi(x)^T F$, leading to 
$$\textstyle \sum_{x\in\X} f(x)p(x) = \tr \Sigma_p F$$ with $\Sigma_p =  \sum_{x\in\X} \varphi(x)\varphi(x)^T p(x)$ the \emph{moment matrix}.
With this in mind, we present in this section KL approximations based on moment matrices.

\subsection{Moment matrix}
For $x \in \X$ and  $I \subset \{0,1\}^d$, we define a \emph{feature vector} indexed by $I$ as $$\varphi_I(x) = (x^\alpha)_{\alpha \in I},$$ where $x^\alpha = \prod_i x_i^{\alpha_i}$. We will often take $I_0 = \{\alpha, \sum_i \alpha_i \leq 1\}= \{0,e_1,\dots,e_d\}$, where $e_i = (\1_{j=i})_{j \in \{1,\dots,d\}}$, for which we denote $\varphi_0(x) = \varphi_{I_0}(x) = (1,x) = (1,x_1,\dots,x_d)$. The $x^\alpha$ are called \emph{monomials} or \emph{features}. We denote $n$ the cardinal of the set $I$. We have $\Vert\varphi(x)\Vert^2 = \sum_{i\in I} \op x^{\alpha_i}\cp^2 = n$. We index the entries of $\varphi(x)$ either by the $\alpha$'s, $\varphi(x)_\alpha = x^\alpha$, or by $\{1,\dots,n\}$.

For a positive measure $p$ on $\X$, we define the \emph{moment matrix} $\Sigma_p$ by:
$$
\Sigma_p = \sum_{x\in\X} \varphi(x)\varphi(x)^Tp(x) = \E_p[\varphi(x)\varphi(x)^T].
$$

\sloppy As $\varphi(x)\varphi(x)^T \in  \S_n^+$ and $p$ is a non-negative measure, we have $\Sigma_p \in \S_n^+$. Moreover, we have $\tr \Sigma = \sum_{x\in\X}\Vert\varphi(x)\Vert^2 p(x) = n$.

As for the vector $\varphi(x)$, the entries of the matrix $\Sigma$ can be indexed either by $i,j \in \{1,\dots,n\}$, or by $\alpha,\beta \in I$. Moreover, we have:
\begin{equation}
\label{eq:entries_sigma}
\begin{array}{rl}
(\Sigma_p)_{\alpha,\beta} &= \E_p[\varphi(x)_\alpha \varphi(x)_\beta] \\
&= \E_p[\prod_i x_i^{\alpha_i+\beta_i}] = \E_p[x^{\alpha\triangle\beta}],   
\end{array}
\end{equation}
where $\triangle$ is the XOR or symmetric difference, $\alpha\triangle\beta = (\1_{\alpha_i \neq \beta_i})_{i=1,\dots,d}$.

We denote $\V$ the linear span of the $\op\varphi(x)\varphi(x)^T\cp_{x\in\X}$, with the following description:
\begin{equation}
\label{eq:V}
\V = \{ M : |I| \times |I| \text{ matrix such that } M_{\alpha,\beta} = M_{\alpha',\beta'} \text{ for all } \alpha \vartriangle\beta =  \alpha'\vartriangle\beta'\},
\end{equation}
where the inclusion $\subset$ comes directly from (\ref{eq:entries_sigma}) and the reserve inclusion $\supset$ comes from the fact that the family of functions $(x\mapsto x^\alpha)_{\alpha \in \{0,1\}^d}$ forms a basis of $\R^\X$, the space of functions from $\X$ to $\R$. For $q$ the uniform measure on $\X$, the random variable $x_1,\dots,x_d$ are independent, hence $\E_q[x^\alpha] = \prod_i\E_p[x_i^{\alpha_i}]$. $\E_q[x_i^{\alpha_i}]$ is equal to $0$ if $\alpha_i = 1$ and $1$ otherwise. With equation (\ref{eq:entries_sigma}), it leads to $\Sigma_q = I_n$.

\subsection{Quantum lower bound on the KL}
We want to define a relaxation of the KL based on moment matrices. Hence it will be defined on $\S_n^+$.

For $A,B \in \S_n^+$, the quantum relaxation of the KL \parencite{matsumotoNewQuantumVersion2018,bachSumofSquaresRelaxationsInformation2024} is :
$$D^{\QT}_V(A,B) = \tr V B^{1/2}h(B^{-1/2}AB^{-1/2})B^{1/2},$$
where $h(t) = t\log t - t +1$ defines a spectral function on $\S_n^+$ and $V \in \S_n^+$ is such that $\varphi(x)^T V \varphi(x) \leq 1$.

This quantity, among a series of relaxation of the KL, is thoroughly studied by \citet{bachSumofSquaresRelaxationsInformation2024}\footnote{Note that more information theory  properties of a different but very similar quantity were also studied in \citet{bachInformationTheoryKernel2023}. Moreover, statistical and geometric properties of the quantum KL have been studied recently by \citet{chazalStatisticalGeometricalProperties2024}.}, from which we learn that $D^{\QT}_V$ is jointly convex in $A$ and $B$, $D^{\QT}_V(A,B) \geq 0$. Moreover, it is a lower bound on the KL: for $p,q \in \P(\X)$, $D^{\QT}_V(\Sigma_p\Vert \Sigma_q) \leq D(p\Vert q)$.

On the boolean hypercube, for any $x\in\X$, $\Vert \varphi(x) \Vert^2 = \sum_{\alpha\in I} \op x^\alpha\cp^2 = n$, therefore $V = I/n$ verifies $V \succcurlyeq 0$ and $\forall x \in \X, \varphi(x)^T V \varphi(x) \leq 1$.\footnote{Instead of taking a fixed matrix $V$, we proposed a method to optimize the bound over a set of such matrices in appendix \ref{annexe:metric_learning}.} For $V = I/n$, we write:
\begin{equation}
\label{eq:KL_QT}
D^{\QT}(A,B) = D^{\QT}_{I/n}(A,B) = \frac{1}{n} \op\tr A\log B^{-1}A -\tr A + \tr B\cp.
\end{equation}
Taking $A = \Sigma_p$ for some probability $p$ and $B = \Sigma_q = I$ for $q$ the uniform distribution, we define:
$$
H(\Sigma_p) = D^{\QT}(\Sigma_p,\Sigma_q) = \frac{1}{n}\tr\Sigma_p\log \Sigma_p,
$$
which corresponds to the von Neumann entropy (see for example \citet{arakiEntropyInequalities1970}).

\section{TRACTABLE UPPER BOUND ON THE LOG-PARTITION\texorpdfstring{\\}{ }FUNCTION}

\label{sct:logp}
\subsection{Upper bound on the log-partition function}

We start from the Donsker-Varadhan identity. To be more general (and make a link with optimization problems later), a temperature parameter $\ep >0$ is added by taking $f\leftarrow f/\ep$ and multiplying the equality by $\ep$:
\begin{eqnarray}
\notag \!\!\!\!\!\Phi_\ep(f) &  \!\!\!= \!\!\!& \textstyle  \ep\log\sum_{x\in\X} e^{f(x)/\ep}q(x) \\
\label{eq:donsker}  & \!\!\!=\!\!\! &  \textstyle \sup_{p\in \P(\X)} \sum_{x\in\X} f(x)p(x) -\ep D(p\Vert q).
\end{eqnarray}
We then get an upper bound on $\Phi_\ep(f)$ by replacing $D(p\Vert q)$ by its lower-bound $D^{\text{QT}}(\Sigma_p\Vert\Sigma_q)$:
\begin{equation}
\label{eq:relax1}
\sup_{p\in P(\X)} \sum_{x\in\X} f(x)p(x) - \ep D^{\text{QT}}(\Sigma_p\Vert\Sigma_q).
\end{equation}
To define our first quantum relaxation of the log-partition function (see section \ref{sct:impr} for more refined relaxations), we take the feature vector $\varphi(x) = \varphi_0(x) = (1,x)$, for which we have $f(x) = \varphi(x)^TF\varphi(x)$ with $F \in \S_n$ defined by $F_{1,i+1} = \theta_{i}$ and $F_{i+1,j+1} = F_{j+1,i+1} = \theta_{ij}/2$. Then, for $p\in\P(\X)$, $\sum_{x\in\X} f(x)p(x) = \tr \Sigma_p F.$ Equation (\ref{eq:relax1}) then becomes:
\begin{equation}
\label{eq:relax2}
\sup_{p\in P(\X), \Sigma = \Sigma_p} \tr F\Sigma - \ep D^{\text{QT}}(\Sigma\Vert\Sigma_q).
\end{equation}
The set $$\K = \{\Sigma = \Sigma_p, p\in P(\X)\}$$ is too complex to lead to a tractable relaxation, therefore we need to find a proper outer approximation.
For $\Sigma\in\K$, there exists a probability $p$ such that $\Sigma = \Sigma_p = \sum_{x\in\X}\varphi(x)\varphi(x)^T p(x)$.
In particular, we have $\Sigma \succcurlyeq 0$, $\Sigma \in \V$ and $\tr \Sigma = n$.
Hence the choice of the outer-approximation $$\K' = \{ \Sigma \succcurlyeq 0, \tr(\Sigma) = n, \Sigma \in \V\}.$$
From (\ref{eq:relax2}), we can then define our quantum relaxation of the log-partition function:
$$
a_\ep(F) = \sup_{\Sigma \in \K'}  \tr \Sigma F -\ep D^{\QT}(\Sigma,\Sigma_q) ,
$$
which verifies $a_\ep(F) \geq \Phi_\ep(f)$.

Using the expression of $D^{\QT}(\Sigma,\Sigma_q)$ given by (\ref{eq:KL_QT}), we get:
$$
a_\ep(F) = \sup_{\Sigma \in \K'}  \tr \Sigma F - \frac{\ep}{n} \op\tr \Sigma\log\Sigma -\tr \Sigma + n\cp,
$$
and because of the constraint $\tr(\Sigma) = n$, it is also equivalent to:
$$
a_\ep(F) = \sup_{\Sigma \in \K'}  \tr \Sigma F - \frac{\ep}{n}\tr \Sigma\log \Sigma.
$$

Assuming that the feature vector is of size polynomial in $d$, this relaxation is tractable as it amounts to manipulating matrices of size polynomial in $d$, and solving a convex optimization problem, which can be done in polynomial time.

\subsection{Computing the bound}
We used Chambolle-Pock algorithm \citep{chambolleFirstOrderPrimalDualAlgorithm2011} to solve the problem defining the quantum relaxation. This algorithm solves a problem of the form:
\begin{equation}
\label{eq:chambolle_primal}
\min_{x \in X} F(x) + G(x),
\end{equation}
where $X$ is a finite-dimensional real vector space with inner product $\langle \cdot,\cdot \rangle $, $G:X \rightarrow[0,+\infty]$ and $F:X \rightarrow[0,+\infty]$ are proper, convex, lower-semicontinuous functions.\footnote{For more details on the convexity results used in this section, see for example \citet{bauschkeConvexAnalysisMonotone2011}.}
Note that this primal problem is equivalent to its dual:
\begin{equation}
\label{eq:chambolle_dual}
\max_{y \in X} -(G^*(-y) + F^*(y)).
\end{equation}
where $G^*:X \rightarrow[0,+\infty]$, $F^*:X \rightarrow[0,+\infty]$ are the Fenchel conjugates of $F$ and $G$.

Here we take $X = \S_n$ and:
\begin{align*}
G : x &= \Sigma \mapsto \frac{\ep}{n}\op\tr\Sigma\log\Sigma -\tr\Sigma + n\cp -\tr \Sigma F\\
F : x &= \Sigma \mapsto \iota_{\V\cap H}(\Sigma),
\end{align*}
with $H = \{\Sigma, \tr\Sigma = n\}$ an affine hyperplane, $\V$ is defined by (\ref{eq:V}) and $\iota_{\V\cap H}: \S_n \rightarrow \{0,+\infty\}$ the indicator function of $\V\cap H$.

The proximal operators are defined, for $\tau > 0$ and $\sigma >0$, by:
\begin{align*}
(I + &\tau\partial F)^{-1}(\Sigma_0)\\
&= \argmin_\Sigma \op F(\Sigma) + \frac{\Vert \Sigma-\Sigma_0\Vert^2}{2\tau} \cp \\
&= \text{proj}_{\V\cap H}(\Sigma_0)\\
&= \text{proj}_\V(\Sigma_0) + \op 1 - \frac{\tr(\text{proj}_\V(\Sigma_0))}{n}\cp I,
\end{align*}
with $\text{proj}_{\V\cap H}$ and $\text{proj}_\V$ the orthogonal projections on the affine spaces $\V\cap H$ and $\V$. We get $(I + \sigma\partial F^*)^{-1}$ through Moreau's identity:
\begin{align*}
(I + &\sigma\partial F^*)^{-1}(\Sigma_0) = \Sigma_0 - \sigma\op I + (1/\sigma)\partial F\cp^{-1}\op\Sigma/\sigma\cp \\
&= \Sigma_0 - \text{proj}_\V(\Sigma_0) - \op \sigma - \frac{\tr(\text{proj}_\V(\Sigma_0))}{n}\cp I.
\end{align*}
And for $G$:
$$
(I + \tau\partial G)^{-1} = \argmin_\Sigma \op G(\Sigma) + \frac{\Vert \Sigma-\Sigma_0\Vert^2}{2\tau} \cp,
$$
the first order condition can be rewritten:
$$
\log \Sigma + \frac{n}{\ep\tau}\Sigma = \frac{n}{\ep}\op F + \frac{\Sigma_0}{\tau}\cp.
$$
Defining the real function $g_\lambda: t \in \R^{++} \mapsto \log t + \frac{t}{\lambda} \in \R$, we can express this first order condition with a spectral function
$
g_{\frac{\tau\ep}{n}}(\Sigma) = \frac{n}{\ep}\op F + \frac{\Sigma_0}{\tau}\cp,
$
hence the solution:
$$
(I + \tau\partial G)^{-1} = g_{\frac{\tau\ep}{n}}^{-1}\op \frac{n}{\ep}\op F + \frac{\Sigma_0}{\tau}\cp\cp.
$$
Moreover, $g_\lambda^{-1}(t) = \lambda\Omega(t - \log\lambda)$ where $\Omega$ is the Wright Omega function, for which we have fast and reliable implementation \citep{lawrenceAlgorithm917Complex2012}.

We also compute the following expressions for the Fenchel conjugates of $F$ and $G$:
\begin{align*}
F^*(Y) &= \tr Y + \iota_{\V^* + \R \cdot I}(Y),\\
G^*(Y) &= \frac{\ep}{n}\tr\op\exp\op\frac{n}{\ep}(Y+F)\cp\cp.
\end{align*}
The algorithm requires parameters $\tau,\sigma > 0$, $\theta \in [0,1]$ and initial point $x_0 \in X$, the updates are defined in (Alg. \ref{alg:chambolle}). We take parameter values $\theta = 1$, $\tau\sigma <1$, for which Theorem 1 of \citet{chambolleFirstOrderPrimalDualAlgorithm2011} guaranties convergence of the iterates.
We experimentally found that it was reasonable to take fixed values of $\tau$ and $\sigma$, which are $\tau = 3$ and $\sigma = 0.3$.

\begin{algorithm}
\caption{Chambolle-Pock algorithm}
\label{alg:chambolle}
\textbf{Initialization:} Choose $\tau,\sigma > 0$, $\theta \in [0,1]$ $x_0,y_0 \in X$ and set $\bar{x}_0 =x_0$

\textbf{Iterations $(n \geq 0)$:} Update $x_n$, $y_n$, $\bar{x}_n$ as follows:
$$
\left\{
    \begin{array}{ll}
        y_{n+1} & = (I + \sigma\partial F^*)^{-1}(y_n + \sigma \bar{x}_n) \\
        x_{n+1} & = (I + \tau\partial G)^{-1}(x_n - \tau y_{n+1}) \\
        \bar{x}_{n+1} & = x_{n+1} + \theta (x_{n+1}-x_n)
    \end{array}
\right.
$$
\end{algorithm}

Using the primal problem (\ref{eq:chambolle_primal}) and the dual (\ref{eq:chambolle_dual}), one can compute a duality gap and define a stopping criterion guarantying some tolerance $\text{tol} > 0$. While the dual points given by the iterates are always feasible, it is not the case for the primal points. We construct a feasible point by first projecting an iterate $\Sigma$ onto $\V\cap H$, then adding the identity matrix to ensure that eigenvalues are non-negative:
$$
\Sigma_\text{feasible}= (1-u)\cdot\text{proj}_{\V\cap H}(\Sigma) + u\cdot I,
$$
with
$
u = 
\left\{
    \begin{array}{ll}
        -\frac{\lambda}{1-\lambda} & \text{if } \lambda = \lambda_{\min} (\text{proj}_{\V\cap H}(\Sigma)) < 0 \\
        0 & \text{otherwise.}
    \end{array}
\right.
$

\subsection{Computational complexity}

Our algorithm involves manipulating matrices of size $n\times n$. We do not need to keep any history of the iterates, hence a space complexity of $O(n^2)$. Assuming that the size $n$ of the feature vector is $O(d)$ ($n=d+1$ for $\varphi(x) = (1,x)$), we get a space complexity $O(d^2)$.

The projection onto the space $\V$ (\ref{eq:V}) can be computed in $O(n^2)$, because we only need to average entries that are equal for matrices in $\V$.
All other steps can be cast by summing and multiplying matrices by a scalar $O(n^2)$, computing the eigenvalue decomposition of a matrix $O(n^3)$, applying a scalar function to a vector of eigenvalues $O(n)$, and multiplying matrices $O(n^3)$.
Hence each step of Chambolle-Pock algorithm as a time complexity in $O(n^3)$.

\citet{chambolleFirstOrderPrimalDualAlgorithm2011} give a theoretical bound on the number of iterations to reach a tolerance of $\delta$ in $O(\frac{1}{\delta})$, with a constant that can depend on the objective function, hence in $n$ and model parameters~$\theta$.

Empirically (see appendix \ref{annexe:convergence_CP}), for feature vector $\varphi(x) = (1,x)$, we found out that we rather had linear convergence in $\delta$, i.e., in $O(\log(\frac{1}{\delta}))$, and that the constant was linear in $d$ for reasonable choices of models parameters. Hence a total time complexity in $O(\log(\frac{1}{\delta})d^4)$.

For a more general feature vector, we do not observe this linear convergence in $\delta$, but rather the theoretical rate in $O(\frac{1}{\delta})$, with a constant in $O(n^2)$, leading to a time complexity in $O(\frac{1}{\delta}n^5)$.

\section{IMPROVING THE BOUND}

\label{sct:impr}
\subsection{Tightness}
\label{sct:impr:tightness}
The ``smallest'' feature vector we can take in order to be sure to be able to write $\sum_{x\in\X} f(x)p(x) = \tr \Sigma_p F$ is $\varphi_0(x) = (1,x)$. Could adding more features make the bound tighter ?

For the feature vector $\varphi(x) = (x^\alpha)_{\alpha \in \{0,1\}^d}$, i.e., taking all possible features, the bound $D^{\QT}(\Sigma_p\Vert\Sigma_q) \leq D(p\Vert q)$ is tight.
Indeed, the vectors $\varphi(x)/2^d$ form an orthogonal basis of $\R^{2^d}$ as for $x,x' \in \X$:
$$
\varphi(x)^T\varphi(x') = \sum_{\alpha \in \{0,1\}^d} x^\alpha x'^\alpha = \prod_i (1 + x_ix_i').
$$
Therefore, the eigenvectors and eigenvalues of $\Sigma_p$ and $\Sigma_q$ are exactly the $\varphi(x)/2^d$ and respectively $2^dp(x)$ and $2^dq(x)$.
It follows that:
\begin{multline*}
D^{\QT}(\Sigma_p,\Sigma_q) = \frac{1}{2^d} \tr\op \Sigma_p\log \Sigma_q^{-1}\Sigma_p - \Sigma_p +  \Sigma_q\cp \\
= \frac{1}{2^d} \bigg(\sum_{x\in\X} 2^dp(x) \log\frac{2^dp(x)}{2^dq(x)} - 2^dq(x) + 2^dq(x)\bigg) 
= \sum_{x\in\X} p(x) \log\frac{p(x)}{q(x)} = D(p\Vert q).
\end{multline*}
However, as the number of features $n = 2^d$ is exponential in $d$, it is not tractable.

\subsection{Hierarchies}
\label{sct:impr:hierarchies}
What about just adding some extra features? Adding extra features defines hierarchies of relaxations, classical in sum-of-squares optimization \citep{lasserreMomentsPositivePolynomials2010}. 

Suppose that a feature vector $\varphi$ defines a quantum relaxation of the log-partition function, with some matrix $F$ such that $f(x) = \varphi(x)^T F \varphi(x)$ as above. If this feature vector is embedded in a larger feature vector $\td\varphi=(\varphi,\varphi_+)$, one can write $f(x) = \td\varphi(x)^T \td F \td \varphi(x)$ with 
$$\td F = \begin{pmatrix} F & 0 \\ 0 & 0 \end{pmatrix} \in \S_{\td n},$$
leading to a relaxation of the log-partition function, defined as an optimization problem over $\td\K  = \{ \td\Sigma \succcurlyeq 0, \tr(\td\Sigma) = \td n, \Sigma \in \td\V\}.$ A matrix $\td \Sigma \in \S_{\td n}$ can be written 
$$\td\Sigma = \begin{pmatrix}
    \Sigma & \Sigma_+\\
    \Sigma_+^T & \Sigma_{++}
\end{pmatrix},$$
with $\Sigma \in \S_{n}$. Having $\td \Sigma \in \td\K'$ implies that $\Sigma \in \K'$, meaning that adding extra feature, like in classical SoS, leads to additional constraints on the optimization domain and an improved value of the upper-bound on the log partition function.

\subsection{Greedy algorithm}
\label{sct:impr:greedy}
Adding additional features can make the bound tighter, however it makes the problem more computationally complex. We explore the idea of adding additional features one by one with a greedy algorithm similar to orthogonal matching pursuit \citep{mallatWaveletTourSignal2009}.

From a feature vector $\varphi_I(x) = (x^\alpha)_{\alpha\in I}$, we compute the set of features at ``distance one'' from $I$: $\td I =\{\alpha\vartriangle e_i, \forall i \in \{1,\dots,d\},\forall\alpha\in I\}\backslash I$, where $e_i = (\1_{j=i})_{j \in \{1,\dots,d\}}$ and $\vartriangle$ is the XOR or symmetric difference. We then select the feature $x^{\hat\alpha} \in \td I$ such that the value of $\td a_\ep(F)$ computed with feature vector $\td\varphi(x) = (x^\alpha)_{\alpha\in I\cup\{\hat\alpha\}}$ is the smallest. The procedure is summarized in Alg.~\ref{alg:greedy}.

Starting from feature vector $\varphi_0 = (1, x)$, we select $k$ features with this procedure. As $|\td I| \leq d|I|$, we compute the value of $a_\ep(F)$ at most $O(k\times d\times(d+k))$ times with feature vectors of size at most $d+k+1$, therefore this algorithm remains polynomial in $d$.

In practice, to select the best $\hat\alpha$ we don't need a high precision on the value of the $\td a_\ep(F)$'s, so we used a higher tolerance during the selection selection process, and then a low tolerance to compute the final value once the features have been chosen.

\begin{algorithm}
\caption{Greedy selection of features}
\label{alg:greedy}
\textbf{Initialization:} $F$ matrix of parameters , $k$ number of features to select.

\textbf{Set:} $I = I_0 = \{0,e_1,\dots,e_d\}$ and $\varphi(x) = \varphi_I(x) = (1,x_1,\dots,x_d)$.

\textbf{For $i = 1, \dots, k$:}

\quad-- Compute $\td I =\{\alpha\vartriangle e_i, \forall i \in \{1,\dots,d\},\forall\alpha\in I\}\backslash I$.

\quad-- For $\hat \alpha \in \td I$: compute  $a(\hat \alpha) = \td a_\ep(F)$ for feature vector $\td\varphi(x) = (x^\alpha)_{\alpha\in I\cup\{\hat\alpha\}}$.

\quad-- Select $\hat{\alpha_0} = \argmin_{\hat \alpha} a(\hat \alpha)$.

\quad-- Update $I~\leftarrow~I~\cup~\{\hat{\alpha_0}\}$.
\end{algorithm}

\section{LOW TEMPERATURE LIMIT}

As $D(p\Vert q)$ and $D^\QT(\Sigma\Vert\Sigma_q)$ are non-negative, $\Phi_\ep(f)$ and $a_\ep(F)$ are non-decreasing with respect to $\ep$. What are the limit of these quantities when $\ep \rightarrow 0$?

From the dual formulation (\ref{eq:donsker}), as $D(p\Vert q)$ is finite for all $p \in \P(\X)$ and $\P(\X)$ is bounded, it follows that:
$$
\Phi_\ep(f) \underset{\ep \rightarrow 0}{\longrightarrow} \sup_{p \in \P(\X)} \sum_{x\in\X} f(x)p(x) = \max_{x\in\X} f(x).
$$
Moreover,  $D^\QT(\Sigma\Vert \Sigma_q)$ is finite for all $\Sigma \in \K'$ with $\K'$ bounded, so we also have:
$$
a_\ep(F) \underset{\ep \rightarrow 0}{\longrightarrow} a_0^1(F) = \sup_{\Sigma \in \K'}  \tr \Sigma F.
$$
$a_0^1(F)$  corresponds to classical tractable SOS relaxation of the optimization problem $\max_{x\in\X} f(x)$ \citep{lasserreMomentsPositivePolynomials2010}. Here, in the special case of the Boolean hypercube, these relaxations are linked to the famous semidefinite relaxations of the MAXCUT problem \citep{goemansImprovedApproximationAlgorithms1995} and nonconvex quadratic optimization \citep{nesterovSemidefiniteRelaxationNonconvex1998}.

When the graph underlying the MRF is a tree, this relaxation is tight as soon as we take the feature vector $\varphi(x) = (\varphi_0(x),\varphi_+(x))$, where $\varphi_+(x) = (x_ix_j)_{ij\in E}$. See appendix \ref{annexe:opt_special} for more details.

\section{EXPERIMENTS}

\label{sct:exp}
\subsection{Competing algorithms}
The state-of-the-art algorithms for the problem of computing upper-bounds on the log-partition function for the Ising model are the log-determinant relaxation from \citet{jordanSemidefiniteRelaxationsApproximate2003} and tree-reweighted message passing (TRW) from \citet{wainwrightNewClassUpper2005a}. We give a brief overview of these methods in appendix \ref{annexe:comp}.

We implemented our algorithms, as well as the competing algorithms, in Python, writing the optimization schemes using NumPy for our algorithms and TRW, and using CVXPY \citep{diamond2016cvxpy} for the log-determinant. We chose to prioritize simplicity and reusability of the code over fully optimized code. Therefore, we do not include an extensive numerical comparison of runtimes as it would only be scientifically sound if all methods were implemented using the same framework (e.g., with first-order methods).

\subsection{Set-up}
\paragraph{Parameters.} To evaluate the relaxations, we need to choose the value of the vector of the parameters $\theta$ in $f(x) = \sum_i \theta_i x_i + \sum_{ij} \theta_{ij} x_i x_j$.

First, we take $\theta \sim \N(0,I)$, and we call this setting \emph{Gaussian parameters}.

We also take the setting of \citet{jordanSemidefiniteRelaxationsApproximate2003}, which we call \emph{log-det parameters}. It corresponds to $\theta_i \sim \U(-0.25,+0.25)$ independently, and distinguishes three cases for the couplings terms: $\theta_{ij} \sim \U(0,2w)$ for \emph{attractive coupling}, $\theta_{ij} \sim \U(-w,w)$ for \emph{mixed coupling} and $\theta_{ij} \sim \U(-2w,0)$ for \emph{repulsive coupling}, where $w \geq 0$ is the \emph{coupling strength}.

\paragraph{Graph.} We also need to choose the graph $G = (V,E)$ underling the MRF. We take $V = \{1,\dots,d\}$ and distinguish three cases for the edges $E$:
\vspace*{-.3cm}
\begin{itemize}
\item independent variable: $E = \varnothing$, i.e., $\theta_{ij} = 0$ for all $i,j$.
\vspace*{-.2cm}
\item tree: $G = (V,E)$ is a tree, i.e., $\theta_{ij} \neq 0$ only on edges of this tree
\vspace*{-.2cm}
\item complete graph: $E = \{(i,j) \text{ for all } i,j \}$, i.e., all $\theta_{ij} \neq 0$ .
\end{itemize}

\paragraph{Metrics.} To evaluation the error in the upper-bound on the log-partition function compare to the true value, we follow \citet{wainwrightNewClassUpper2005a} and define the \emph{normalized error in bound} as
$$
\mathit{error\_bound} = ( \mathit{bound} - \mathit{true\_value})/d.
$$
As in \citet{jordanSemidefiniteRelaxationsApproximate2003,wainwrightNewClassUpper2005a}, the approximated marginals $\hat{p}$ is evaluated with a $l_1$ error,
$$\mathit{l1\_error} = \textstyle \frac{1}{d}\sum_{i\in\{1,\dots,d\}} | \hat{p}(x_i = 1) - p(x_i = 1)|.$$

\subsection{Hierarchies}
We start by comparing our different relaxations for different level of hierarchies. We take Gaussian parameters, and $d=5$, so that we can take up to all possible features. We plot the value of our relaxations versus the size of the feature vector, starting with $\varphi_0(x) = (1,x)$ of size $6$. For the quantum relaxation, we take features $\alpha$ with increasing numbers of non-zero coordinate, while for the quantum relaxation we select features with the greedy procedure presented in section \ref{sct:impr:greedy}.

\begin{figure*}[p]
    \centering
	\subfloat[][independent variables]{
    \includegraphics[width=.32\linewidth]{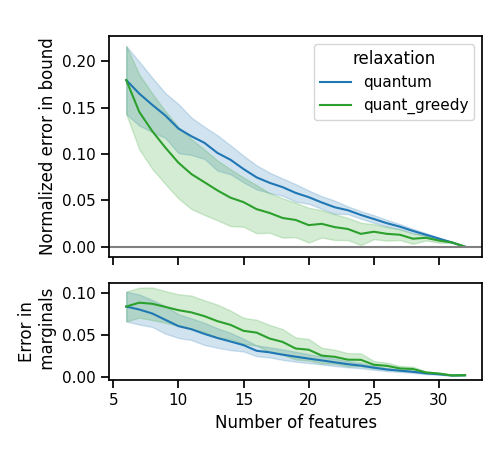}
    }
    \subfloat[][tree]{
    \includegraphics[width=.32\linewidth]{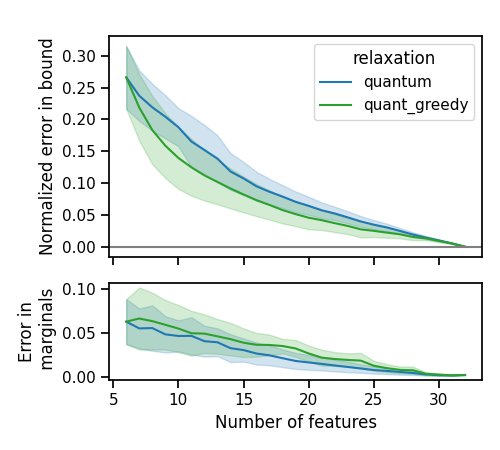}
    }
    \subfloat[][complete graph]{
    \includegraphics[width=.32\linewidth]{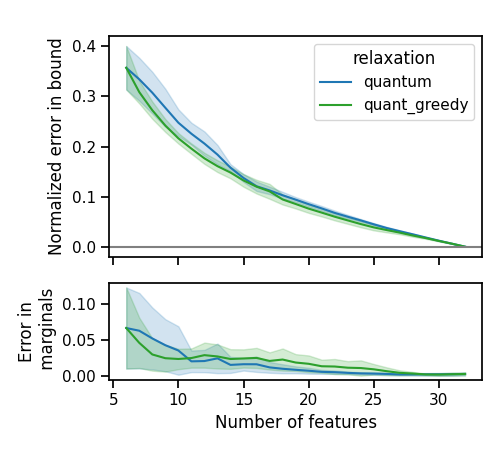}
    }
	\caption{Comparison of the different quantum relaxations of the KL divergence. $d=5$, Gaussian parameters. x-axis corresponds to the number of features in the feature vector. Mean over 10 draws of parameters, $\pm1$ standard deviation.}
	\label{fig:logp_1}
\end{figure*}

From Fig. \ref{fig:logp_1}, we see that the greedy procedure allows to select interesting features quickly, and allows to reduce the gap with less features than adding features in arbitrary order, in particular for independent variable or MRF on a tree, while the gains for the complete graph are less pronounced.

\subsection{Comparison with competing algorithms}
We compare the relaxations on the log-det parameters for varying coupling strength.\footnote{Comparisons for other choices of parameters are done in appendix \ref{annexe:TRW_param} and appendix \ref{annexe:gaussian_param}.} We limit ourselves to $d=5$ and coupling strength $w < 0.5$, as we observed convergence issues with TRW algorithm. For the greedy algorithm, we take $k=3$ extra features.\footnote{We take other values of $k$ in appendix \ref{annexe:greedy}, showing even stronger bounds at the expense of more computations.}

\begin{figure*}[p]
    \centering
	\subfloat[][attractive coupling]{
    \includegraphics[width=.32\linewidth]{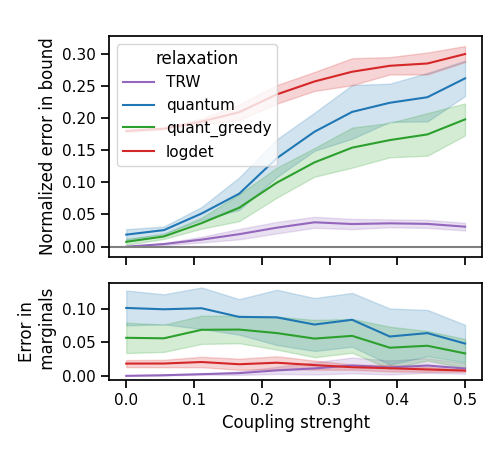}
    }
    \subfloat[][mixed coupling]{
    \includegraphics[width=.32\linewidth]{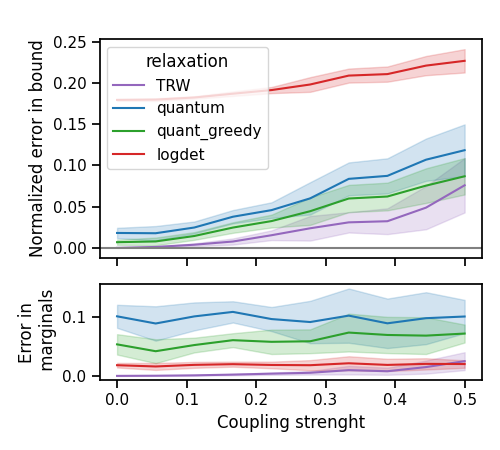}
    }
    \subfloat[][repulsive coupling]{
    \includegraphics[width=.32\linewidth]{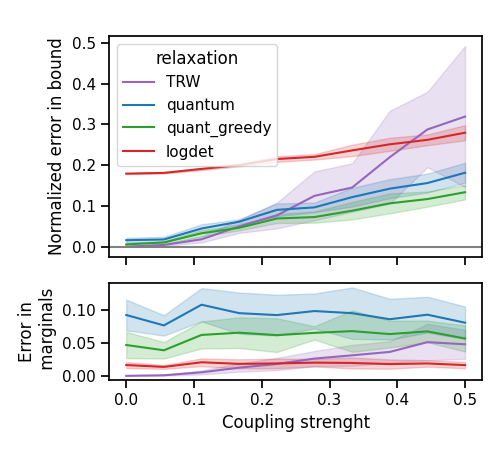}
    }
    \caption{Comparison with competing algorithm for log-det parameters and varying coupling strength. $d = 5$, complete graph. Mean over 10 draws of parameters, $\pm1$ standard deviation.}
	\label{fig:comp_2}
\end{figure*}

In Fig. \ref{fig:comp_2}, we use the log-det parameters on the complete graph.
Our relaxations give better bound on the log-partition function than the log-determinant relaxation for all coupling strengths and interaction types.
They are competitive with the TRW relaxation for low coupling strength.
For attractive coupling and high coupling strength, TRW is better.
However, for mixed couplings, the greedy relaxation stays competitive with TRW even at higher coupling strength, and for repulsive coupling, both relaxations become better than the TRW relaxation at high coupling strength. For the $\mathit{l1\_error}$, our relaxations exhibit higher error than both the TRW and log-determinant relaxations.

We also carried out additional experiments with a larger value of $d$, $d = 16$. As we observed convergence issues with the message passing scheme of the TRW relaxation, leading to imprecise results and causing the optimization of $\rho$  to fail, we decided to use a fixed uniform value for $\rho$, $\rho_{ij} = 2/d$ (such that $\sum_{ij}\rho_{ij} = d-1$), and to stop message passing after a maximum of $10000$ iterations.\footnote{Note moreover that in the absence of convergence, there is no more guaranty that the output is an upper-bound on the true value.}

\begin{figure*}[p]
    \centering
	\subfloat[][attractive coupling]{
    \includegraphics[width=.32\linewidth]{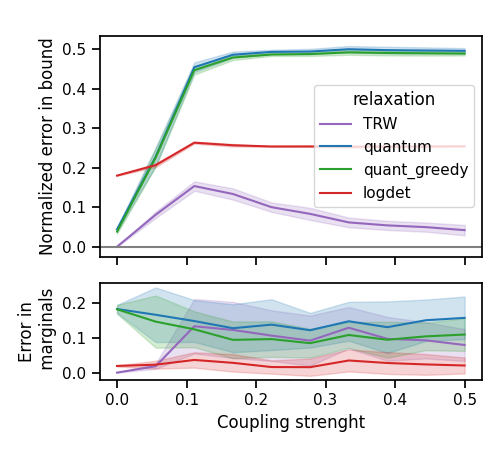}
    }
    \subfloat[][mixed coupling]{
    \includegraphics[width=.32\linewidth]{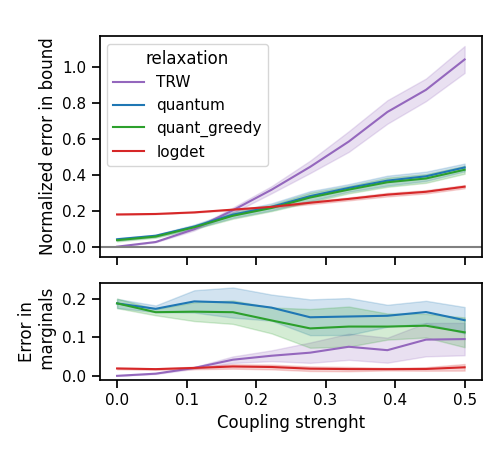}
    }
    \subfloat[][repulsive coupling]{
    \includegraphics[width=.32\linewidth]{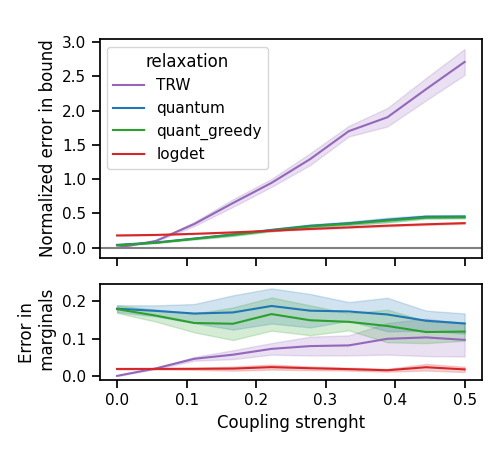}
    }
    \caption{Comparison with competing algorithm for log-det parameters and varying coupling strength. $d = 16$, complete graph. Mean over 10 draws of parameters, $\pm1$ standard deviation.}
	\label{fig:comp_rho_fixed_2}
\end{figure*}

In Fig. \ref{fig:comp_rho_fixed_2}, we use the log-det parameters on the complete graph.  For attractive coupling, our relaxations are not as good as competing algorithms as soon as the coupling strength increases, but stay competitive with the log-determinant relaxation for mixed coupling and repulsive coupling, while the performance of the TRW relaxation deteriorates.

\section{CONCLUSION}

In this work, we developed from \citet{bachInformationTheoryKernel2023,bachSumofSquaresRelaxationsInformation2024}, and studied relaxations of the KL divergence and the log-partition function on the Boolean hypercube $\X = \{-1,1\}^d$. The relaxations we propose are competitive with the best competing algorithms for the same problem, and even better in some settings. 

They are several avenues to be explored to extend this work. 
As these methods rely on a feature vector, they can be adapted on other spaces $\X$, paving the way to a universal framework for variational inference in a wide variety of setting beyond the Boolean hypercube.
We could also try to improve our algorithms by using recent improvements in SDP and SoS optimization \citep[see, e.g.,][]{huangSolvingSDPFaster2022,jiangFasterInteriorPointMethod2023}.
It would be interesting to get theoretical guaranties on our upper-bounds of the log-partition function, we could try using recent convergence results of SoS optimization for the hypercube \citep{baldiDegreeBoundsPutinars2024} or take inspiration from works on other variational inference methods \citep[see, e.g.,][]{risteskiHowCalculatePartition2016,jainMeanfieldApproximationConvex2019}.
Moreover, there are strong links to be made with sum-of-squares optimization, allowing to tackle complex ML pipelines involving both inference and optimization.
Finally, the algorithms we developed do not take into account the special structure of the graphical model. Future work could be dedicated to investigate how our methods can gain from incorporating more knowledge on the problem structure.

\subsubsection*{Acknowledgements}
The authors thank the reviewers for their feedback. This work has received support from the French government, managed by the National Research Agency, under the France 2030 program with the reference ``PR[AI]RIE-PSAI'' (ANR-23-IACL-0008).

\FloatBarrier

\printbibliography

\clearpage

\appendix
\appendixpage

\section{COMPETING ALGORITHMS}

\label{annexe:comp}
\subsection{The log-determinant relaxation}
\citet{jordanSemidefiniteRelaxationsApproximate2003} use the same variational inference principle as presented above, but with the following convex bound on entropy:
$$
H(p) \leq \frac{1}{2}\log\det \left[\Sigma_p + \frac{1}{3} \begin{pmatrix}I_d & 0 \\ 0 & 0 \end{pmatrix}\right] + \frac{d}{2}\log \left(\frac{\pi e}{2}\right),
$$
where $\Sigma_p$ is the $d+1\times d+1$ moment matrix for the feature vector $\varphi_0(x) = (1,x)$. It corresponds to bounding by the entropy of a Gaussian vector with similar covariance matrix. For the outer approximation of $\K$ , they also take $\K'$, but they also propose to add additional linear constraints to have consistency with pairwise marginals. More precisely, it consists in enforcing the following inequalities, for all $i,j \in \{1,\dots,n\}$ and for all $a,b \in \{-1,1\}$:
$$
1 +a\mu_i + b\mu_j +ab\mu_{ij} \geq 0,
$$
where $\mu_i = \Sigma_{1,i+1}$ (corresponding to $\E_p[x_i]$ for $\Sigma_p$) and $\mu_{ij} = \Sigma_{i+1,j+1}$ (corresponding to $\E_p[x_ix_j]$ for $\Sigma_p$).

Note that this upper bound on entropy is not tight, even for independent variables, by construction, as we upper bound the entropy of a non Gaussian distribution by the one of a Gaussian with matching covariance matrix.
Even for the uniform distribution, it is not tight. Indeed, for $p$ the uniform distribution :
$$
H(p) = d\log2 < d \log\left(\sqrt{\frac{2\pi e}{3}}\right)
$$
with $\sqrt{\frac{2\pi e}{3}} \approx 2.386$.

\subsection{Tree-reweighted Message Passing (TRW)}

\citet{wainwrightNewClassUpper2005a} reformulate the problem Eq. \ref{eq:donsker} in terms of moment vector $\mu = (\mu_i,\mu_{ij})$, where $\mu_i = \E_p[x_i]$ for $i\in \{1,\dots,n\}$ and $\mu_{ij} = \E_p[x_ix_j]$, for $i,j \in \{1,\dots,n\}$.

To construct a bound on entropy, they use the spanning tree polytope, which is the set of vectors $\rho = (\rho_{ij}) \in \R^{d(d-1)/2}$, also called edge appearance probabilities, that verify:
$$
\rho_{ij} = \E_\mathbb{P}[\1_{(i,j)\in T}]
$$
where $\mathbb{P}$ is a probability on spanning trees $T$ of the MRF graph $G = (V,E)$ and $\1_{(i,j)\in T} \in \{0,1\}$ indicates whether $(i,j)$ is an edge of $T$. Then for such a vector $\rho$, they introduce the following bound on entropy:
$$
H(\mu) \leq \sum_{s\in V} H_s(\mu_s) - \sum_{(s,t)\in E}\rho_{st}I_{st}(\mu_{st}),
$$
and deduce an upper bound on the log-partition function as done above. For fixed $\rho$, they solve the variational problem with a message-passing scheme, and then they optimize over $\rho$ with the conditional gradient algorithm \citep{frankAlgorithmQuadraticProgramming1956}.

\section{TIGHTNESS OF THE RELAXATION FOR ZERO TEMPERATURE AND SPECIAL STRUCTURE IN THE PARAMETERS}

\label{annexe:opt_special}
\subsection{Independent variables}
Taking $f(x) = \sum_i \theta_i x_i$ leads to the variable $x_1,\dots,x_d$ being independent under the corresponding probability $p$.

Already with feature vector $\varphi_0(x) = (1,x)$, the constraints $\Sigma \in \K'$ implies the positive definiteness of the matrices:
$$
\begin{pmatrix}
1 & \mu_i\\
\mu_i & 1
\end{pmatrix}, 
$$
with $\mu_i = \Sigma_{1,i+1}$. It follows that $\forall i, -1 \leq \mu_i \leq 1$. In particular, there is a probability $p'$ such that $\forall i, \E_{p'}[x_i] = \mu_i$. However, the other entries of $\Sigma$ might not match with $\Sigma_p$.

In the case of optimization $\ep = 0$, this does not matter as the matrix $F$ that parameterizes a probability with independent variables is non-zero only for the entries $F_{1,i+1} = F_{i+1,1}$, and therefore
$$
\sup_{\Sigma \in \K'} \tr(\Sigma F)  = \sup_{p} \tr (\Sigma_p F).
$$

\subsection{Probability that factorizes over a tree}
We now consider an MRF that factorizes over a tree, meaning that $f(x) = \sum_i \theta_i x_i + \sum_{(i,j)\in E}\theta_{ij}x_ix_j$ for some tree $T =(V,E)$.

Now taking feature vector $\varphi(x) = (\varphi_0(x),\varphi_+(x))$, where $\varphi_+(x) = (x_ix_j)_{ij\in E}$, the constraints $\Sigma \in \K'$ implies the positive definiteness of the matrices:
$$
\begin{pmatrix}
1 & \mu_i\\
\mu_i & 1
\end{pmatrix}, 
\begin{pmatrix}
1 & \mu_i & \mu_j & \mu_{ij}\\
\mu_i & 1 & \mu_{ij} & \mu_j\\
\mu_j & \mu_{ij} & 1 & \mu_i\\
\mu_{ij} & \mu_j & \mu_i & 1
\end{pmatrix}.
$$
with $\mu_i = \Sigma_{1,i+1}$, and $\mu_{ij} =\Sigma_{i+1,j+1}$, hence:
$$\mu_i \in [-1,1],$$
$$\frac{1}{4}[1 + a \mu_i + b \mu_j + a b \mu_{ij}] \geq 0, \forall (i,j) \in E, \forall a,b \in \{-1,1\}.$$
The $\mu_i$'s and $\mu_{ij}$'s are pairwise consistent marginals over the edges and vertices of a tree, which implies that there is a probability $p'$ over $\X$ such that $\E_{p'}[x_i] = \mu_i$ and $\E_{p'}[x_i x_j] = \mu_{ij}$ (see for example Proposition 2.1. of \citet{wainwrightGraphicalModelsExponential2007}).
As previously, this implies tightness of the bounds in the case of optimization $\ep =0$, but not in general.

\section{METRIC LEARNING}

\label{annexe:metric_learning}
\subsection{Quantum relaxation with metric learning}
The quantum relaxation of the KL is given by:
$$D^{\QT}_V(A,B) = \tr V B^{1/2}h(B^{-1/2}AB^{-1/2})B^{1/2},$$
with $V$ such that $\varphi(x)^* V \varphi(x) \leq 1$.
Can we optimize the bound with respect to $V$? In fact it corresponds to optimizing over linear transformation of the feature vector $\td\varphi(x) = T\varphi(x)$ through $V = T^T T$ \citep{bachSumofSquaresRelaxationsInformation2024}.

We limit ourselves to what we call \emph{diagonal metric learning}, with $V = \diag(\eta), \eta \succcurlyeq 0, \sum_i\eta_i = 1$, for which $\varphi(x)^T V \varphi(x) = \sum_{\alpha\in I} \eta_\alpha \op x^\alpha\cp^2 = 1$. 
It leads to define:
\begin{multline*}
D^{\QT}_{\diag}(A,B) = \sup_{\eta \succcurlyeq 0, \sum_i\eta_i = 1} \tr \diag(\eta)(A\log B^{-1}A - A + B)
=\max(\diag(A\log B^{-1}A - A + B)),  
\end{multline*}
and
\begin{equation}
\label{eq:relax_metric_learning}
a_\ep^{\diag}(F) = \sup_{\Sigma \in \K'} \tr\Sigma F - \ep\max(\diag (\Sigma\log(\Sigma) - \Sigma + I).
\end{equation}

The convex problem defining $a_\ep^{\diag}(F)$ can be optimized with Kelley's method \citep{kelleyjr.CuttingPlaneMethodSolving1960}. For more details, see the following subsection. 

We experimentally found that the gain on the bound of the log-partition function were not very important albeit increased computational complexity. Therefore, it seems more interesting to simply take the quantum relaxation and add features with the greedy algorithm. 

\subsection{Optimization algorithm}
\subsubsection{Overview of the Kelley's method}
Kelley's method \citep{kelleyjr.CuttingPlaneMethodSolving1960} solves optimization problems of the form:
\begin{equation*}
\begin{aligned}
    &\min & & f(x)\\
    &\text{ s.t.} & & x \in \X.
\end{aligned}
\end{equation*}
We assume that we can compute subgradients of the convex function $f$ for any $x\in \X$, and that we can solve linear optimization problems over $\X$ (in particular $\X$ must be compact).

From iterates $x_0,\dots,x_t \in \X$, convexity of $f$ gives that $ \forall i \in \{0,\dots,t\}, \forall x \in X, f(x)\geq f(x_i) + g_i\cdot(x-x_i)$, where $g_i \in \partial f(x_i)$, hence $\forall x \in X, f(x)\geq h_t(x)$ where $h_t(x) = \sup_{i=0,\dots,t} f(x_i) + g_i\cdot(x-x_i)$.
We construct next iterate $x_{t+1}$ by solving the linear program $\min_{x\in\X} h_t(x)$, which gives a lower-bound on the optimal value.

Denoting $\hat{f}_{t} = \min_{x\in\X} h_{t-1}(x)$ and $\td{f}_{t} = \min_{i=0,\dots,t} f(x_i)$, $\hat{f}_{t}$ is a lower-bound on the optimal value, increasing with $t$ and $\td{f}_{t}$ is an upper-bound on the optimal value, decreasing with $t$, hence the natural stopping criterion $\td{f}_{t} - \hat{f}_{t} < \text{tol}$.

\subsubsection{Use in quantum relaxation with metric learning}
To solve (\ref{eq:relax_metric_learning}) with Kelley's method, we take $f : x = \Sigma \mapsto \tr\Sigma F - \ep\max(\diag (\Sigma\log(\Sigma) - \Sigma + I)$ and constraints $\Sigma \in \K'$.

Solving the inner linear program and dealing with the constraints can be done with  any optimization framework. In this work, we used CVXPY \citep{diamond2016cvxpy}. We also need to find the expression of a subgradient of the function $g: \Sigma \mapsto \max(\diag (\Sigma\log(\Sigma) - \Sigma + I) = \max_i g_i(\Sigma)$, with  $g_i(\Sigma) = e_i^T(\Sigma\log(\Sigma) - \Sigma + I) e_i = \tr(e_ie_i^T(\Sigma\log(\Sigma) - \Sigma + I))$ and $(e_i)_i$ is the canonical basis of $\R^n$.

Using result form section \ref{annexe:diff_spectral}, we have:
$$
\nabla g_i (\Sigma) = \sum_{kl} \frac{h(\lambda_i)- h(\lambda_j)}{\lambda_i-\lambda_j} u_l^Te_ie_i^Tu_k u_lu_k^T = \sum_{kl}(u_k)_i(u_l)_i\frac{h(\lambda_k)- h(\lambda_l)}{\lambda_k-\lambda_l} u_lu_k^T.
$$
with $\Sigma = \lambda_i u_i u_i^T$ the orthogonal decomposition of sigma and $h: t \mapsto t\log t - t + 1$.

Then, with theorem 18.5 of \citet{bauschkeConvexAnalysisMonotone2011}, it follows that $\partial g(\Sigma) = \conv(\nabla g_i (\Sigma))_{i\in \td I}$, the convex envelope of $(\nabla g_i (\Sigma))_{i\in \td I}$, where $\td I = \{i : h(\Sigma)_{ii} = \max(\diag (h(\Sigma)))\}$.

\section{DIFFERENTIATING SPECTRAL FUNCTIONS}

\label{annexe:diff_spectral}
To solve the optimization problems of \ref{annexe:metric_learning}, we need to compute gradient of a function involving terms of the form $A \mapsto f(A)$. To achieve this, we use a trick based on Cauchy’s residue formula due to \citet{katoPerturbationTheoryLinear1966}.

Let $A$ be a symmetric matrix and $f: \mathbb{C} \rightarrow \mathbb{C}$ a holomorphic function.  We write $A = \sum_i \lambda_i u_i u_i^T$. With Cauchy’s residue formula, we get:
\begin{align*}
f(A) &= \sum_i f(\lambda_i) u_i u_i^T\\
&= \frac{1}{2i\pi}\oint_\gamma f(z) (zI - A)^{-1} dz,
\end{align*}
where $\gamma$ is a contour enclosing all eigenvalues.
For a small perturbation $\Delta$, the eigenvalues of $A + \Delta$ remain enclosed by $\gamma$, and:
\begin{align*}
f(A+\Delta) - f(A) &= \frac{1}{2i\pi}\oint_\gamma f(z) \ob(zI - A - \Delta)^{-1} - (zI - A)^{-1}\cb dz \\
&= \frac{1}{2i\pi}\oint_\gamma f(z) (zI - A)^{-1}\Delta(zI - A - \Delta)^{-1} dz \\
&= \frac{1}{2i\pi}\oint_\gamma f(z) (zI - A)^{-1}\Delta(zI - A)^{-1} dz + o(\Delta)\\
&= \frac{1}{2i\pi}\oint_\gamma \sum_{ij} \frac{f(z)}{(z-\lambda_i)(z-\lambda_j)} u_iu_i^T\Delta u_ju_j^T dz  + o(\Delta)\\
&= \sum_{ij} \frac{f(\lambda_i)- f(\lambda_j)}{\lambda_i-\lambda_j} u_iu_i^T\Delta u_ju_j^T + o(\Delta),
\end{align*}
as $\frac{1}{(z-\lambda_i)(z-\lambda_j)} = \frac{1}{\lambda_i-\lambda_j} \op \frac{1}{z-\lambda_i} - \frac{1}{z-\lambda_j} \cp \text{ for } \lambda_i \neq \lambda_j$, and denoting $\frac{f(\lambda_i)- f(\lambda_j)}{\lambda_i-\lambda_j}=f'(\lambda_i)$ if $\lambda_i = \lambda_j$. This leads to the following expression of the differential:
$$
D_f(A): \Delta \mapsto \sum_{ij} \frac{f(\lambda_i)- f(\lambda_j)}{\lambda_i-\lambda_j} u_iu_i^T\Delta u_ju_j^T.
$$

For $g: A \mapsto \tr B f(A)$, where $B \in M_n$ the chain rule gives
\begin{align*}
D_g(A): \Delta \mapsto &\tr (B D_f(A)(\Delta)) \\
& = \tr \op\op\sum_{ij} \frac{f(\lambda_i)- f(\lambda_j)}{\lambda_i-\lambda_j} u_j^TBu_i u_ju_i^T \cp \Delta\cp,
\end{align*}
hence,
$$
\nabla g (A) = \sum_{ij} \frac{f(\lambda_i)- f(\lambda_j)}{\lambda_i-\lambda_j} u_j^TBu_i u_ju_i^T.
$$
In particular, for $B = I$, the formula becomes $\nabla g (A) = f'(A)$.

\section{ADDITIONAL EXPERIMENTS}

\subsection{Hierarchies on the KL divergence}
Here we compare our different relaxations on the KL divergence for different level of hierarchies. We take Gaussian parameters, and $d=5$, so that we can take up to all possible features. We plot the value of our relaxations of the KL divergence versus the size of the feature vector, starting with $\varphi_0(x) = (1,x)$ of size $6$. For the quantum relaxation, we take features $\alpha$ with increasing numbers of non-zero coordinate, while for the quantum relaxation we select features with the greedy procedure presented in section \ref{sct:impr:greedy}.

\begin{figure*}
    \centering
	\subfloat[][independent variables]{
    \includegraphics[width=.32\linewidth]{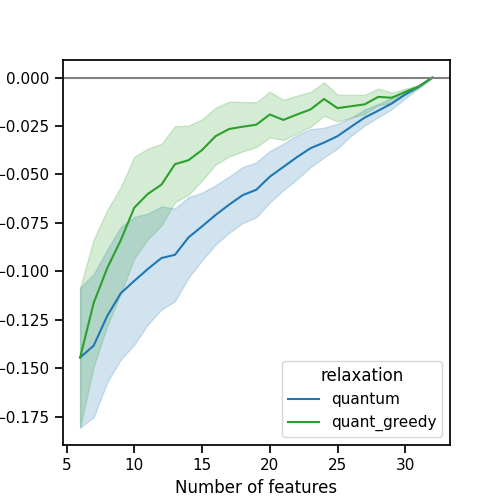}
    }
    \subfloat[][tree]{
    \includegraphics[width=.32\linewidth]{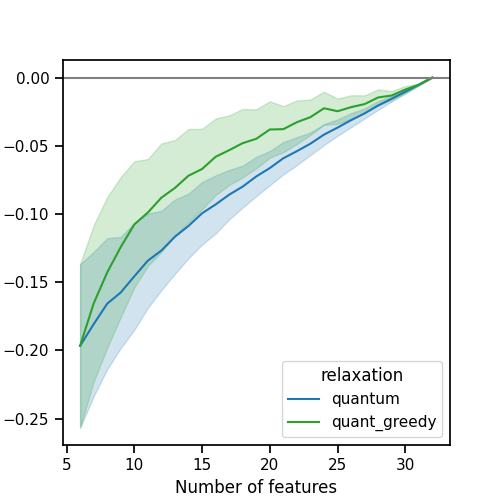}
    }
    \subfloat[][complete graph]{
    \includegraphics[width=.32\linewidth]{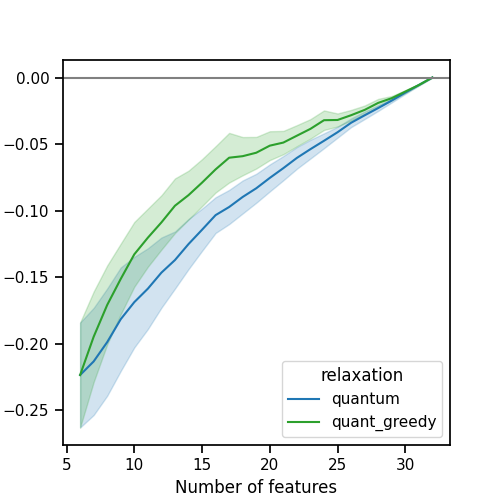}
    }
	\caption{Comparison of the different quantum relaxations of the KL divergence. $d=5$, Gaussian parameters. x-axis corresponds to the number of features in the feature vector. Mean over 10 draws of parameters, $\pm1$ standard deviation.}
	\label{fig:KL_1}
\end{figure*}

From Fig. \ref{fig:KL_1}, we see that the greedy procedure allows to select interesting features quickly, and allows to reduce the gap with less features than adding features in arbitrary order. For the feature vector $\varphi(x) = (x^\alpha)_{\alpha\in \{0,1\}^d}$, the relaxation is tight.

\subsection{TRW parameters}
\label{annexe:TRW_param}
We make  additional experiments in the setting of \citet{wainwrightNewClassUpper2005a}, we call \emph{TRW parameters}. It corresponds to $\theta_i \sim \U(-0.05,+0.05)$ independently, and distinguishes two cases for the couplings terms: $\theta_{ij} \sim \U(0,w)$ for \emph{attractive coupling} and $\theta_{ij} \sim \U(-w,w)$ for \emph{mixed coupling}, where $w$ is called the \emph{coupling strength}.

\begin{figure}
    \centering
	\subfloat[][attractive coupling]{
    \includegraphics[width=.45\linewidth]{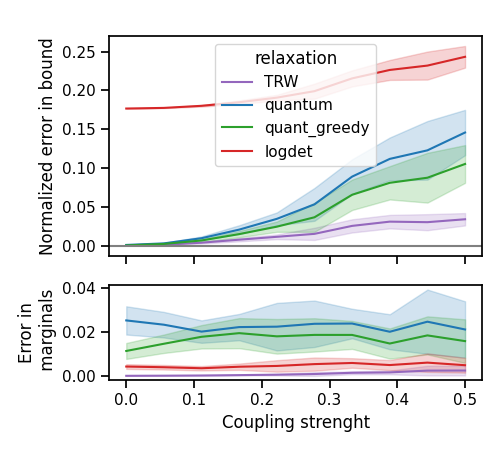}
    }
    \subfloat[][mixed coupling]{
    \includegraphics[width=.45\linewidth]{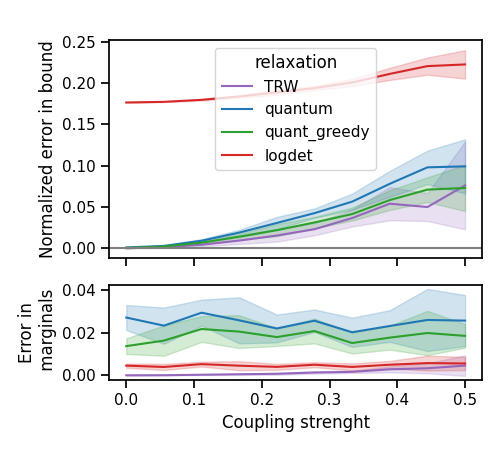}
    }
	\caption{Comparison with competing algorithm for TRW parameters and varying coupling strength. $d=5$, complete graph. Mean over 10 draws of parameters, $\pm1$ standard deviation.}
	\label{fig:comp_1}
\end{figure}

In Fig. \ref{fig:comp_1}, we use the TRW parameters on the complete graph. Our relaxations on the log-partition function are competitive with the TRW relaxation for low coupling strength.
For mixed couplings, the greedy relaxation stays competitive with TRW even at higher coupling strength.
The log-determinant relaxation exhibits a much higher error on the log-partition function. For the $\mathit{l1\_error}$, our relaxations exhibit higher error than both the TRW and log-determinant relaxations.

We also carried out additional experiments with a larger value of $d$, $d = 16$. In this case, for the TRW relaxation, we use a fixed uniform value for $\rho$, ($\rho_{ij} = 2/d$) and stop message passing after a maximum of $10000$ iterations.

\begin{figure}
    \centering
	\subfloat[][attractive coupling]{
    \includegraphics[width=.45\linewidth]{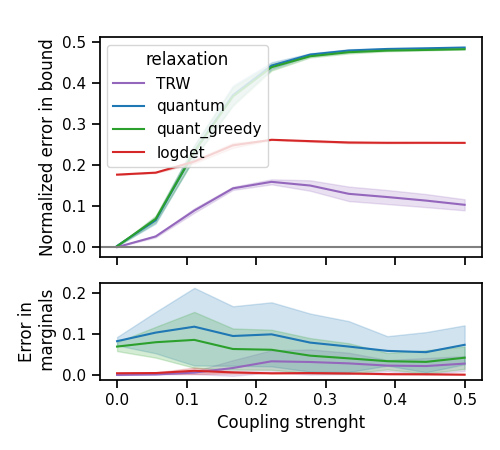}
    }
    \subfloat[][ mixed coupling]{
    \includegraphics[width=.45\linewidth]{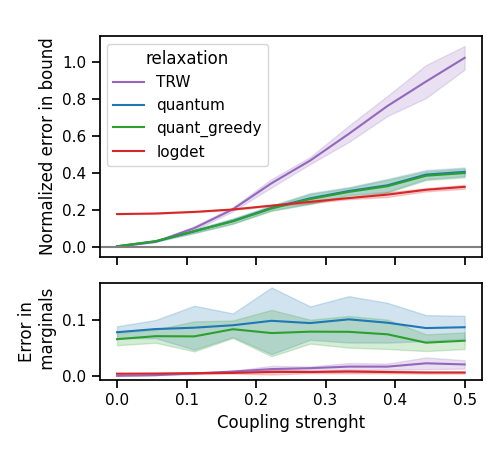}
    }
	\caption{Comparison with competing algorithm for TRW parameters and varying coupling strength. $d = 16$, complete graph. Mean over 10 draws of parameters, $\pm1$ standard deviation.}
	\label{fig:comp_rho_fixed_1}
\end{figure}

In Fig. \ref{fig:comp_rho_fixed_1}, we use the TRW parameters on the complete graph. For attractive coupling, our relaxations are not as good as competing algorithms as soon as the coupling strength increases. However, they stay competitive with the log-determinant relaxation for mixed coupling, while the TRW relaxation becomes significantly worse at high coupling strength. For the $\mathit{l1\_error}$, our relaxations still exhibit higher error than both the TRW and log-determinant relaxations.

\subsection{Gaussian parameters and varying temperature}
\label{annexe:gaussian_param}
We compare with competing algorithms on the Gaussian parameters for varying temperature. We take $d = 10$, and for the TRW relaxation, we use a fixed uniform value for $\rho$, ($\rho_{ij} = 2/d$) and stop message passing after a maximum of $10000$ iterations.

\begin{figure}
    \centering
	\includegraphics[width=.45\linewidth]{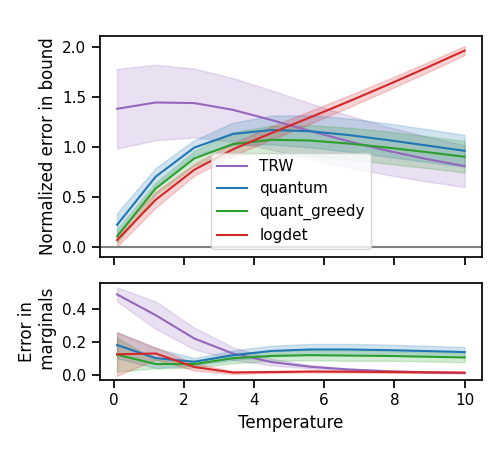}
	\caption{Comparison with competing algorithms for Gaussian parameters and varying temperature. $d=10$, complete graph. Mean over 10 draws of parameters, $\pm1$ standard deviation.}
	\label{fig:comp_3}
\end{figure}

In Fig. \ref{fig:comp_3}, we see that our relaxations are competitive with the log-determinant relaxation at low-temperature, and as the temperature increases it reaches a plateau whereas the error on the log-determinant relaxation grows linearly. They are also better than the TRW relaxation at low temperature and stay competitive as the temperature increase. For the $\mathit{l1\_error}$, our relaxation are better than the log-determinant and TRW relaxation at low temperature, but worse at high temperature.

\subsection{Number of added greedy features}
\label{annexe:greedy}
We're now taking a closer look at the impact of the number of features added by the greedy procedure. To this end, we introduce the normalized relative gain in bound, defined by:
$$
\mathit{gain\_bound} = \frac{\mathit{bound} - \mathit{quantum\_bound}}{d},
$$
where $\mathit{bound}$ is the value of the bound with added features and $ \mathit{quantum\_bound}$ is the bound for the regular quantum relaxation with $\varphi(x) = (1,x)$.

We take $d = 10$, and for the TRW relaxation, use a fixed uniform value for $\rho$, ($\rho_{ij} = 2/d$), and stop message passing after a maximum of $10000$ iterations.

\begin{figure*}
    \centering
	\subfloat[][Normalized gain in bound]{
    \includegraphics[height=6cm]{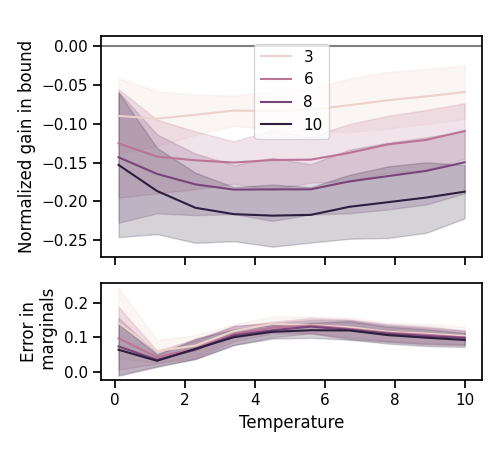}
    }
    \subfloat[][Normalized error in bound]{
    \includegraphics[height=6cm]{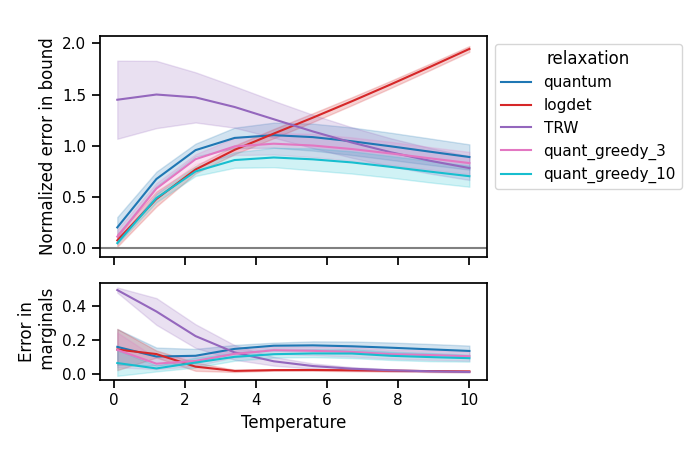}
    }
    \caption{(a) Normalized gain in bound compared to quantum relaxation for different number of features added with the greedy procedure. (b) Comparison with competing algorithms. $d = 10$, complete graph, Gaussian parameters. Mean over 10 draws of parameters, $\pm1$ standard deviation.}
	\label{fig:hierarchies_greedy_1}
\end{figure*}

Experiments with Gaussian parameters are plotted in Fig. \ref{fig:hierarchies_greedy_1}. In plot (a), we observe that additional features increase the normalized relative gain in bound over the whole range of temperature. In plot (b), we see the gains for $k=3$ and $k=10$ when compared with the log-determinant and TRW relaxations. Adding $10$ extras features makes the relaxation better than the two competing algorithms.

\begin{figure*}
    \centering
	\subfloat[][attractive coupling]{
    \includegraphics[width=.32\linewidth]{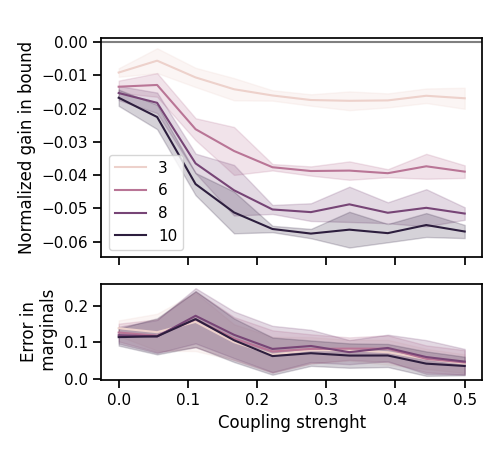}
    }
    \subfloat[][mixed coupling]{
    \includegraphics[width=.32\linewidth]{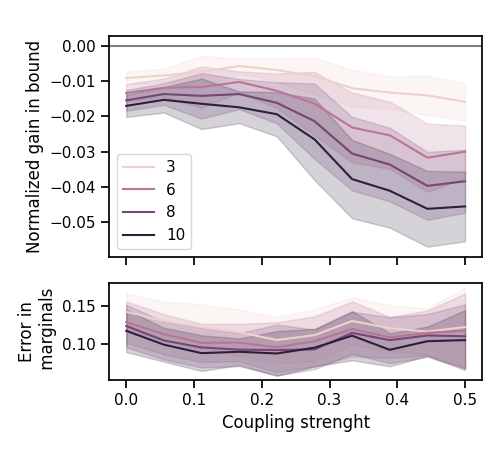}
    }
    \subfloat[][repulsive coupling]{
    \includegraphics[width=.32\linewidth]{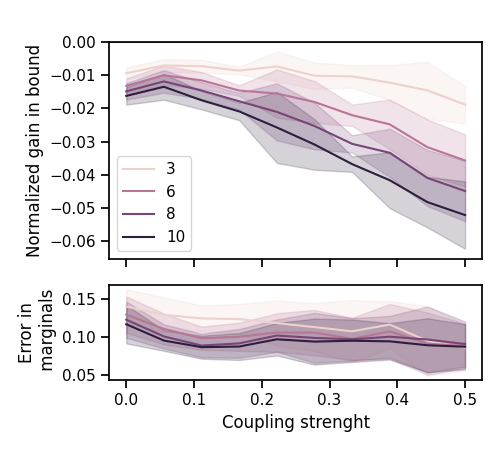}
    }
    \caption{Normalized gain in bound compared to quantum relaxation for different number of features added with the greedy procedure. $d = 10$, complete graph, log-det parameters. Mean over 10 draws of parameters, $\pm1$ standard deviation.}
	\label{fig:hierarchies_greedy_2a}
\end{figure*}

\begin{figure*}
    \centering
	\subfloat[][attractive coupling]{
    \includegraphics[width=.32\linewidth]{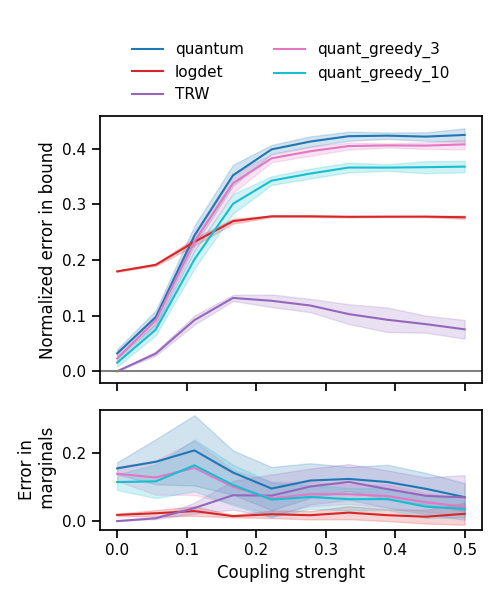}
    }
    \subfloat[][mixed coupling]{
    \includegraphics[width=.32\linewidth]{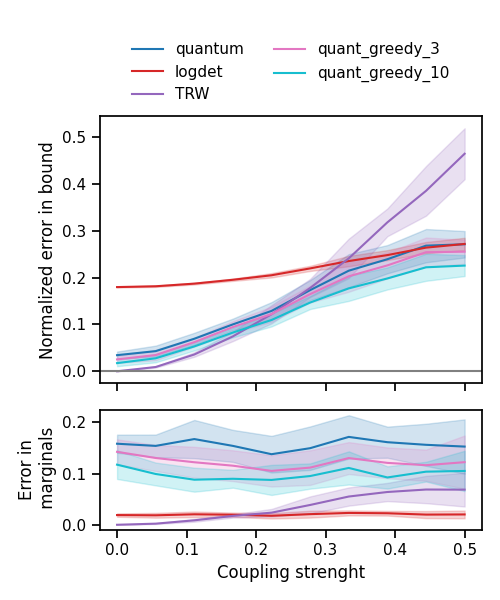}
    }
    \subfloat[][repulsive coupling]{
    \includegraphics[width=.32\linewidth]{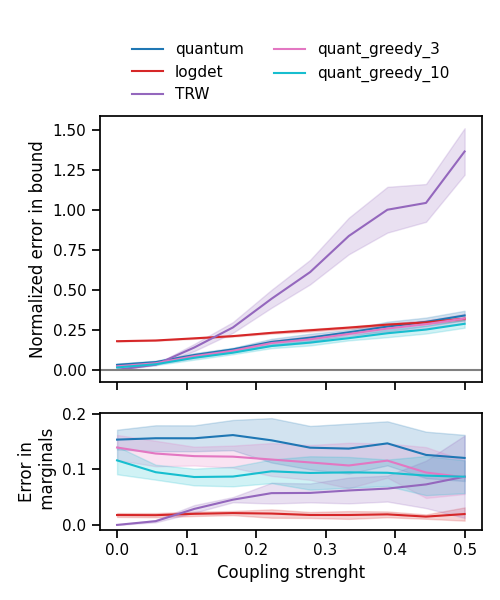}
    }
    \caption{Comparison with competing algorithms. $d = 10$, complete graph, log-det parameters. Mean over 10 draws of parameters, $\pm1$ standard deviation.}
	\label{fig:hierarchies_greedy_2b}
\end{figure*}

Experiments with log-det parameters are plotted in Fig.~\ref{fig:hierarchies_greedy_2a} and Fig.~\ref{fig:hierarchies_greedy_2b}. In Fig.~\ref{fig:hierarchies_greedy_2a}, we observe that additional features increase the normalized relative gain in bound over the whole range of coupling strength, and type of couplings, with a larger gain at high coupling strength. In Fig.~\ref{fig:hierarchies_greedy_2b}, we see the gains for $k=3$ and $k=10$ when compared with the log-determinant and TRW relaxations. This gains allow to beat the log-determinant relaxation at high coupling strength for both mixed coupling and repulsive coupling, and mitigate the gap between our relaxations and competing algorithm for attractive coupling.

\subsection{Higher dimension}
We carried additional experiments with higher value of $d$. As the complexity of the brute force solution growth exponentially fast with $d$, we can't compute the exact value of the log-partition function and marginals. Therefore, we compare our bound to the log-determinant relaxation, by defining:
$$
\mathit{relative\_error\_bound} = \frac{\mathit{bound} - \mathit{bound\_log\_det}}{d}.
$$
When this value is negative, relaxation is better than log-determinant relaxation, when it is positive, relaxation is worse.

We take $d = 30$, and for the TRW relaxation, we use a fixed uniform value for $\rho$, $\rho_{ij} = 2/d$ (such that $\sum_{ij}\rho_{ij} = d-1$), and stop message passing after a maximum of $10000$ iterations.

\begin{figure*}
    \centering
	\subfloat[][attractive coupling]{
    \includegraphics[width=.32\linewidth]{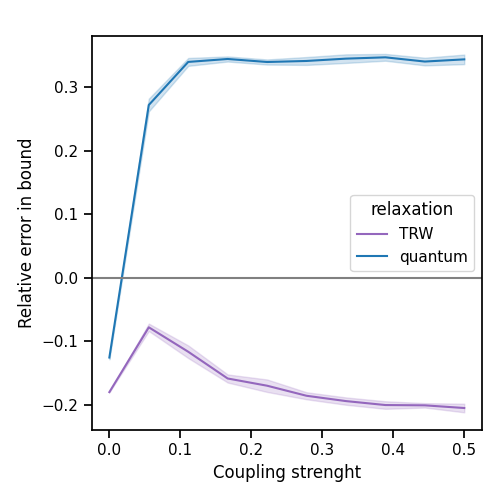}
    }
    \subfloat[][mixed coupling]{
    \includegraphics[width=.32\linewidth]{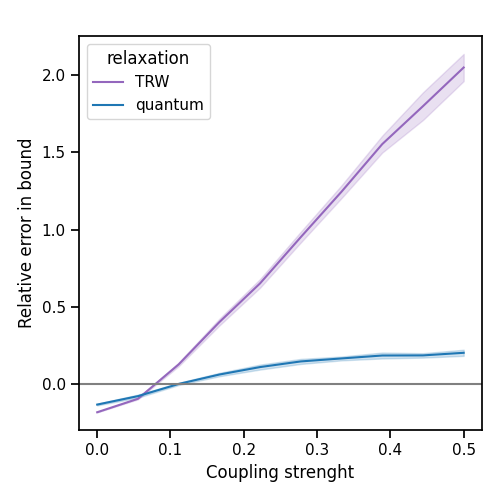}
    }
    \subfloat[][repulsive coupling]{
    \includegraphics[width=.32\linewidth]{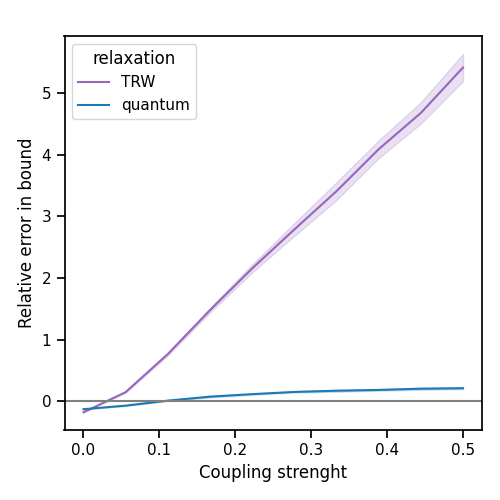}
    }
    \caption{Comparison with competing algorithm for log-det parameters and varying coupling strength. $d = 30$, complete graph. Mean over 10 draws of parameters, $\pm1$ standard deviation.}
	\label{fig:comp_large_d}
\end{figure*}

In Fig. \ref{fig:comp_large_d}, we observe results that are similar to Fig. \ref{fig:comp_rho_fixed_2}. For attractive coupling, our relaxations are not as good as competing algorithms as soon as the coupling strength increase, but stay competitive with the log-determinant relaxation for mixed coupling and repulsive coupling, while the performance of the TRW relaxation deteriorates.

\subsection{Convergence of Chambolle-Pock algorithm}
\label{annexe:convergence_CP}

We studied empirically the number of iterations of Chambolle-Pock algorithm with respect to $d$ and the tolerance on the duality gap $\delta$.

\begin{figure}
    \centering
	\includegraphics[width=.6\linewidth]{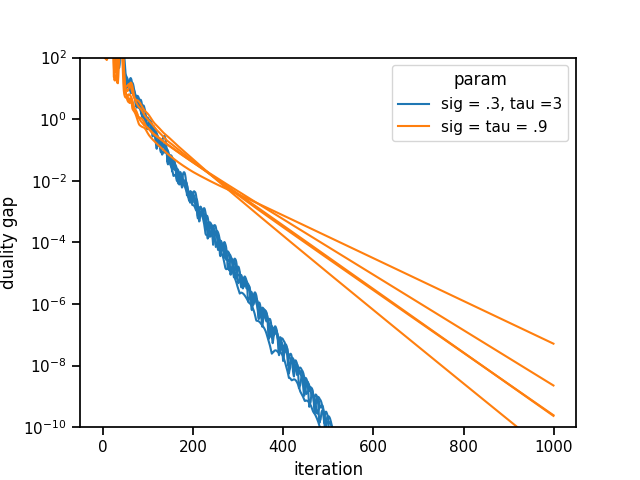}
	\caption{Duality gap along iterations of Chambolle-Pock algorithm, for 5 draws of models parameters and different values of $\tau$ and $\sigma$. $d=50$, Gaussian parameters, complete graph, $\varphi(x) = (1,x)$. y-axis is logarithmic scale.}
	\label{fig:CP_gap_iter}
\end{figure}

For feature vector $\varphi(x) = (1,x)$, in Fig. \ref{fig:CP_gap_iter}, we observe that the duality gap decrease exponentially fast with the number of iterations $t$. Hence the number of iterations is $O(\log(\frac{1}{\delta}))$. 

\begin{figure}
    \centering
	\includegraphics[width=.6\linewidth]{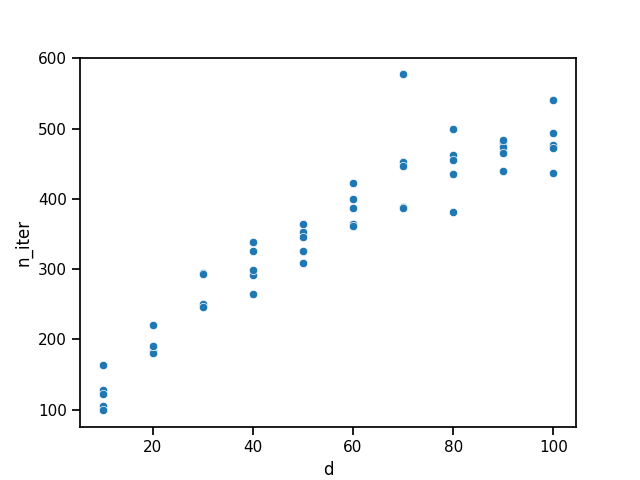}
	\caption{Number of iterations to reach a duality gap $<10^{-8}$, for various values of $d$. $5$ draws of models parameters at each value $d$, Gaussian parameters, complete graph, $\varphi(x) = (1,x)$.}
	\label{fig:CP_iter_d}
\end{figure}

We then study the dependence of the constant with regard to $d$, for $\varphi(x) = (1,x)$. We took random samples of the models parameters according to a normal distribution $\theta \sim \N(0,I)$, such that the scale of the individuals $\theta_i, \theta_{ij}$ does not vary with $d$ : $\theta_i, \theta_{ij} = O(1)$. In Fig. \ref{fig:CP_iter_d}, we observe that the number of iterations to reached a fixed tolerance of $10^{-8}$ grows linearly with $d$.

\begin{figure}
    \centering
	\includegraphics[width=.6\linewidth]{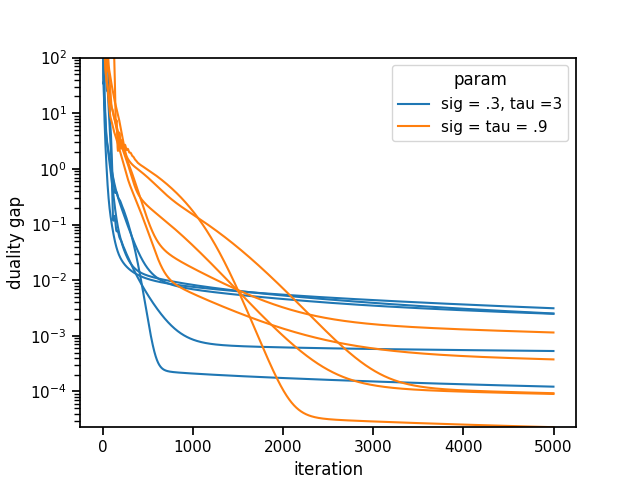}
	\caption{Duality gap along iterations of Chambolle-Pock algorithm, for 5 draws of models parameters and different values of $\tau$ and $\sigma$. $d=50$, Gaussian parameters, complete graph, $\varphi(x) = (1,x)$. y-axis is logarithmic scale.}
	\label{fig:CP_gap_iter_ho}
\end{figure}

\begin{figure}
    \centering
	\includegraphics[width=.6\linewidth]{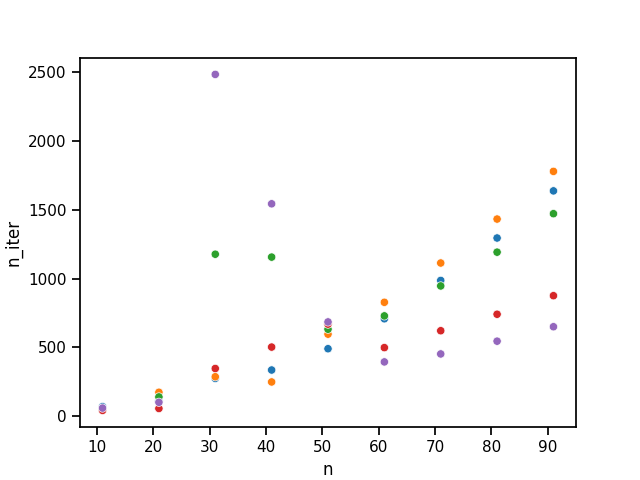}
	\caption{Number of iterations to reach a duality gap $<10^{-2}$, for various values of $n$, with fixed $d$. $5$ draws of models parameters, $d = 10$, Gaussian parameters, complete graph, $\varphi(x) = (1,x)$.}
	\label{fig:CP_iter_n}
\end{figure}

In Fig. \ref{fig:CP_gap_iter_ho}, we fixed $d=10$, and take a feature vector of size $n=50$ (features are chosen in arbitrary order, by increasing degree). For this more general feature vector, we do not observe this linear convergence in $\delta$, and can only hope to get the theoretical rate in $O(\frac{1}{\delta})$. Reaching arbitrary small tolerance becomes much more time consuming than in the case of $\varphi(x) = (1,x)$, but we observe that we can still reach lower precision in reasonable time. In Fig. \ref{fig:CP_iter_n}, we fixed $d=10$, and we observe that the number of iterations to reached a fixed tolerance of $10^{-2}$ for various values of $n$. We observed a general growth in $O(n^2)$, though for some draws and values of $n$, the number of iterations can be much bigger. 

\end{document}